\documentclass{cambridge7A}
  \usepackage{natbib}

  \usepackage{graphicx}

\let\aap=\aaa

\let\aapr=\aarev

\newcommand\an{{\em Astron.\ Nachr.}}
\newcommand\apjl{{\em ApJ (Letters)}}

\let\apjs=\apjsupp
\newcommand\apj{{\em ApJ}}

\newcommand\apss{{\em Ap\&SS}}

\newcommand\mnras{{\em MNRAS}}

\newcommand\nature{{\em Nature}}
\let\nat=\nature

\newcommand\pasj{{\em PASJ}}
\newcommand\pasp{{\em PASP}}

\let\solphys=\solp

\newcommand{\DPobs}{\mbox{$\Delta P_{\rm obs}$}}
\newcommand{\DPg}{\mbox{$\Delta P_{\rm g}$}}
\newcommand{\numax}{\mbox{$\nu_{\rm max}$}}
\newcommand{\Teff}{\mbox{$T_{\rm eff}$}}
\newcommand{\Dnu}{\mbox{$\Delta \nu$}}
\newcommand{\dnu}[1]{\mbox{$\delta \nu_{#1}$}}
\newcommand{\acena}{\mbox{$\alpha$~Cen~A}}
\newcommand{\acenb}{\mbox{$\alpha$~Cen~B}}

\newcommand{\taucet}{\mbox{$\tau$~Cet}}
\newcommand{\bhyi}{\mbox{$\beta$~Hyi}}
\newcommand{\ihor}{\mbox{$i$~Hor}}
\newcommand{\xihya}{\mbox{$\xi$~Hya}}

\newcommand{\zher}{\mbox{$\zeta$~Her}}
\newcommand{\nuind}{\mbox{$\nu$~Ind}}
\newcommand{\muara}{\mbox{$\mu$~Ara}}

\newcommand{\eboo}{\mbox{$\eta$~Boo}}

\newcommand{\muHz}{\mbox{$\mu$Hz}}

\newcommand{\half}{{\textstyle\frac{1}{2}}}

\newcommand{\Msun}{\mbox{$M_{\odot}$}}

\newcommand{\MOST}{{\em MOST\/}}
\newcommand{\WIRE}{{\em WIRE\/}}

\newcommand{\corot}{{\em CoRoT\/}}
\newcommand{\kepler}{{\em Kepler\/}}

\begin{document}

\chapter{Solar-like Oscillations: An Observational Perspective}

 \begin{center}
\vspace*{-15ex}
{\bf Timothy R. Bedding}\\
Sydney Institute for Astronomy (SIfA), School of Physics,
University of Sydney, Australia
\end{center}

\section{What are solar-like oscillations?}

Oscillations in the Sun are excited stochastically by convection.  We use
the term {\em solar-like\/} to refer to oscillations in other stars that
are excited by the same mechanism, even though some of these stars may be
very different from the Sun.  The stochastic nature of the excitation
produces oscillations over a broad range of frequencies, which in the Sun
is about 1 to 4\,mHz (the well-known 5-minute oscillations).  Stellar
oscillations can also be excited via opacity variations (the {\em
heat-engine mechanism}, also called the $\kappa$~mechanism), as seen in
various types of classical pulsating stars (Cepheids, RR Lyraes, Miras,
white dwarfs, $\delta$~Scuti stars,
etc.).  

For a star to show solar-like oscillations it must be cool enough to have a
surface convective zone.  In practice, this means being cooler than the red
edge of the classical instability strip, which includes the lower main
sequence, as well as cool subgiants and even red giants.  Indeed,
solar-like oscillations with periods of hours (and longer) have now been
observed in thousands of G and K giants (see Sec.~\ref{sec.redgiants}).
There is also good evidence that the pulsations in semiregular variables
(M~giants) are solar-like \citep{ChDKM2001,Bed2003,TBK2010}, as perhaps are
those in M supergiants such as Betelgeuse \citep{KSB2006}.  


What about hotter stars?  By definition, solar-like oscillations are
excited stochastically in the outer convection zone.  Given that surface
convection is thought to inhibit operation of the $\kappa$~mechanism
\citep[e.g.,][]{gastine+dintrans2011-quenching}, it seems that solar-like
and heat-engine oscillations would be mutually exclusive.  In fact, both
may operate in a star provided the surface convection layer is thin
\citep{SGH2002,K+M2010} and there is evidence from {\em Kepler} data for
solar-like oscillations in at least one $\delta$~Scuti star
\citep{kepler-antoci++2011}.  There has even been a suggestion that
sub-surface convection in B-type stars could excite solar-like oscillations
\citep{CLB2009,BDN2010}.  There is some evidence from \corot\ observations
that the $\beta$~Cephei star V1449~Aql shows solar-like oscillations
\citep{BSG2009}, although other interpretations are possible
\citep{DBC2009} and {\em Kepler} observations of similar stars have so far
failed to confirm stochastically-excited oscillations \citep{BPDeC2011}.
Meanwhile, \citet{DBA2010} suggested that the O-type star HD~46149 shows
stochastically-excited oscillations, based on \corot\ photometry.

\section{Properties of oscillations}

We now summarise the main properties of solar-like oscillations.  For more
extensive discussion see, for example, \citet{B+G94}, \citet{B+K2003},
\citet{ChD2004}, \citet{Bal2010} and \citet{AChDK2010}.

\subsection{Types of oscillation modes}
\label{sec.modes}

Broadly speaking, there are two types of stellar oscillations.  Pressure
modes ({\em p~modes}) are acoustic waves, for which the restoring force
arises from the pressure gradient.  These are seen in the Sun and sun-like
stars, and also in most of the classical pulsators.  For gravity modes
({\em g~modes}) the restoring force is buoyancy.  The best studied examples
occur in white dwarfs \citep[see][for a recent
review]{althaus++2010-white-dwarfs}, but g~modes are also seen in other
classical pulsators such as $\gamma$~Dor stars, slowly pulsating B~stars
and some sdB stars.  Some stars have been found to have both p~modes and
g~modes, including hybrid $\delta$~Scuti/$\gamma$~Dor stars \citep{GAB2010}
and possibly the Sun \citep{GJM2008}.  Finally, some stars oscillate in
{\em mixed modes}, which have p-mode character in the envelope and g-mode
character in the core.  These are discussed in Secs.~\ref{sec.mixed}
and~\ref{sec.redgiants}.  First, however, we consider p~modes in some
detail because they are most relevant to the topic of solar-like
oscillations.

Mathematically, each p~mode can be described by three integers.  The {\em
radial order\/} ($n$) specifies the number of nodal shells of the standing
wave.  In the Sun, the modes with the highest amplitude have $n$ in the
range 19--22.  The {\em angular degree\/} ($l = 0, 1, 2, \ldots$) specifies
the number of nodal lines at the surface.  Modes with $l=0$ are called {\em
radial\/} modes and those with $l \ge 1$ are {\em non-radial\/} modes.  
In addition, modes with $l=0$ are sometimes called {\em monopole} modes,
while those with $l=1$ are {\em dipole} modes,
those with $l=2$ are {\em quadrupole} modes, 
and those with $l=3$ are {\em octupole} modes.  Since the surface of a
distant star is generally not resolved, cancellation effects mean that only
modes with $l \le 3$ (or perhaps~4) are observable.  

The third integer that specifies a mode is the {\em azimuthal order\/}
($m$), which takes on values from $-l$ to $+l$.  However, the oscillation
frequencies do not depend on~$m$ unless the star is rotating (or spherical
symmetry is broken in some other way, such as by the presence of a magnetic
field).  For a rotating star, the modes with a given $n$ and $l$ are split
in frequency into a multiplet with $2l+1$ components (one for each value
of~$m$).  The frequency separation between these components is proportional
to the rotational frequency of the star and their relative amplitudes
depend on the inclination angle between the stellar rotation axis and the
line of sight.  The study of rotational splitting in solar-like
oscillations, while showing great promise, is still in its infancy and will
not be discussed further in this chapter.


The comments in the preceding two paragraphs also apply to g~modes, but
note that (i)~radial g~modes do not exist (and, therefore, neither do
radial mixed modes), (ii)~$n$ is generally taken to be negative for g~modes
(see Sec.~3.5.2 of \citealt{AChDK2010} for a full discussion), and
(iii)~rotational splitting of g~modes depends on~$l$ \citep[Sec~3.8.4
of][]{AChDK2010}.

\subsection{The frequency spectrum of oscillations}
\label{sec.spectrum}

\begin{figure}
\centerline{\includegraphics[width=\textwidth]{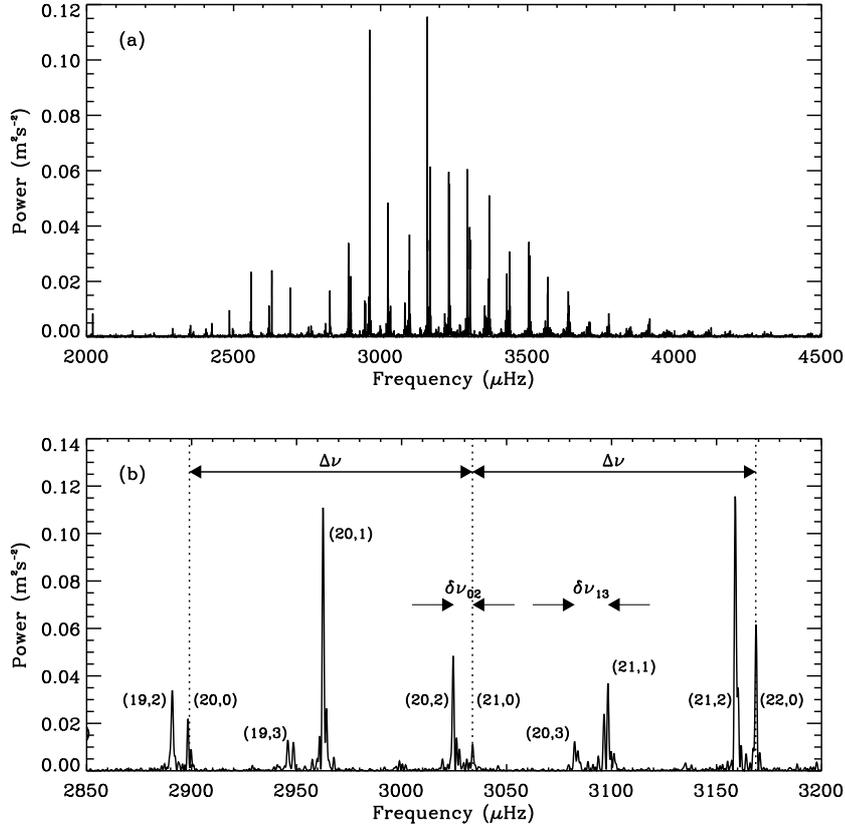}}
\caption{\label{fig.bison.ps} The solar power spectrum from ten days of
disk-integrated velocity measurements with the BiSON instrument
\citep[e.g.,][]{CEI97}.  Panel~(b) shows a close-up labelled with $(n,l)$
values for each mode.  Dotted lines show the radial modes, and the large
and small separations are indicated.  }
\end{figure}

Figure~\ref{fig.bison.ps}a shows the frequency spectrum of p~modes in the
Sun.  The envelope of the peak heights defines the frequency of maximum
power, \numax, which has a value of about 3100\,\muHz\ in the Sun.  The
mode frequencies show a very regular pattern that is characteristic of an
oscillating sphere.  Figure~\ref{fig.bison.ps}b shows a close-up labelled
with the $(n,l)$ values of the modes, which were determined by comparing
with theoretical models.  Note the alternating pattern of large and small
separations as $l$ cycles repeatedly through the values $2, 0, 3, 1, 2, 0,
\ldots$.  The quantity $\Dnu$ is the spacing between consecutive radial
overtones (that is, modes with a given $l$ whose $n$ values differ by 1).
It is called the {\em large frequency separation\/} and has a value of
135\,\muHz\ in the Sun.  To a good approximation, \Dnu\ is proportional to
the square root of the mean stellar density (see Sec.~\ref{sec.scaling}).

In Fig.~\ref{fig.bison.ps}b we see two so-called {\em small frequency
separations}: $\dnu{02}$ is the spacing between close pairs with $l=0$ and
$l=2$ (for which $n$ differs by 1) and $\dnu{13}$ is the spacing between
adjacent modes with $l=1$ and $l=3$ (ditto).  Another small separation can
be defined by noting that the $l=1$ modes do not fall exactly halfway
between $l=0$ modes.  We define $\dnu{01}$ to be the amount by which $l=1$
modes are offset from the midpoint of the $l=0$ modes on either
side.\footnote{One can also define an equivalent quantity, $\dnu{10}$, as
the offset of $l=0$ modes from the midpoint between the surrounding $l=1$
modes, {so that $\dnu{10} = \nu_{n, 0} - \half(\nu_{n-1, 1} +
\nu_{n,1})$.}}  To be explicit, for a given radial order, $n$, these
separations are defined as follows:
\begin{eqnarray}
  \dnu{02} & = & \nu_{n, 0} - \nu_{n-1,2}  \label{eq.dnu02} \\
  \dnu{01} & = & \half(\nu_{n, 0} + \nu_{n+1,0}) - \nu_{n, 1}
       \label{eq.dnu01} \\ 
  \dnu{13} & = & \nu_{n, 1} - \nu_{n-1,3}. \label{eq.dnu13}
\end{eqnarray}
In practice, the oscillation spectrum is not precisely regular and so all
of these spacings vary slightly with frequency.

The $l=4$ modes are weakly visible in Fig.~\ref{fig.bison.ps}b (at 3000 and
3136\,\muHz).  By analogy with $l=2$, we can define the small separation
for these modes as
\begin{equation}
  \dnu{04}  =  \nu_{n, 0} - \nu_{n-2,4}.  \label{eq.dnu04} 
\end{equation}

Finally, we note that each peak in the power spectrum in
Fig.~\ref{fig.bison.ps}b is slightly broadened, which reflects the finite
lifetimes of the modes.  If we define the lifetime~$\tau$ of a damped
oscillation mode to be the time for the amplitude to decay by a factor
of~$e$ then this mode will produce a Lorentzian shape in the power spectrum
with a linewidth (FWHM) of $\Gamma = (\pi\tau)^{-1}$.  The dominant modes
in the Sun have linewidths of 1--2\,\muHz\ and hence lifetimes of 2--4~days
\citep[e.g.,][]{CEI97}.

\subsection{The \'echelle diagram}
\label{sec.echelle}

The \'echelle diagram, first introduced by \citet{GFP83} for global
helioseismology, is used extensively in asteroseismology to display
oscillation frequencies.  It involves dividing the spectrum into equal
segments of length \Dnu\ and stacking them one above the other so that
modes with a given degree align vertically in ridges.  Any departures from
regularity are clearly visible as curvature in the \'echelle diagram.  For
example, variations in the small separations appear as a convergence or
divergence of the corresponding ridges.

\begin{figure}
\centerline{\includegraphics[width=0.6\textwidth]{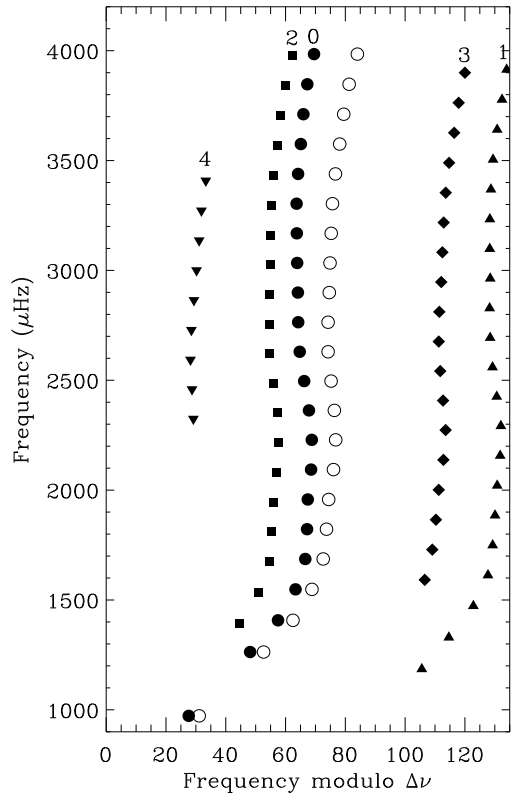}}
\caption{\label{fig.bison.echelle} \'Echelle diagram of the observed
frequencies in the Sun \citep{BCD2009}.  The filled symbols show the
frequencies of modes with $l=0$, 1, 2, 3 and~4, using a large separation of
$\Dnu = 135.0$\,\muHz.  The open symbols show the result of plotting the
$l=0$ frequencies modulo a slightly smaller large separation ($\Dnu =
134.5$\,\muHz).  }
\end{figure}

Figure~\ref{fig.bison.echelle} shows the observed frequencies in the Sun in
\'echelle format for modes with $l \le4$ (filled symbols), using a large
separation of $\Dnu = 135.0$\,\muHz.  Using a slightly smaller value of
\Dnu\ causes the ridges to tilt to the right, as shown by the open symbols
for $l=0$ only.  It also shifts the ridges sideways, which is important if
we are trying to measure the quantity~$\epsilon$ (see
Sec.~\ref{sec.asymptotic}).

Sometimes the ridges wrap around, which has almost happened to the $l=1$
modes at the top of Fig.~\ref{fig.bison.echelle} (filled triangles).  Of
course, the diagram can be made wider by plotting the ridges more than once
(an example is shown in the right panel of Fig.~\ref{fig.etaboo}).  Also
note that one can shift the ridges sideways by subtracting a fixed
reference frequency, which must be kept in mind when reading $\epsilon$
from the diagram (again, see Fig.~\ref{fig.etaboo}).

When making an \'echelle diagram, it is usual to plot $\nu$ versus ($\nu
\bmod \Dnu$), in which case each order slopes upwards slightly (see
Fig.~\ref{fig.bison.echelle}).  However, for greyscale images it can be
preferable to keep the orders horizontal (an example is shown
Fig.~\ref{fig.procyon.echelle}).

\subsection{The asymptotic relation}
\label{sec.asymptotic}

The approximate regularity of the p-mode spectrum means we can express the
mode frequencies in terms of the large and small separations, as follows:
\begin{equation}
  \nu_{n,l} \approx \Dnu (n + \half l + \epsilon) - \dnu{0l}.
        \label{eq.asymptotic}
\end{equation}
Here, $\nu_{n,l}$ is the frequency of the p~mode with radial order $n$ and
angular degree $l$ (and assuming no dependence on~$m$, which means
neglecting any rotational splitting).  By definition, the small separation
\dnu{0l} is zero for $l=0$.  Its values for $l=1$, 2 and 4 have already
been discussed (Eqs.~\ref{eq.dnu02}, \ref{eq.dnu01}, and~\ref{eq.dnu04}).
For $l=3$ we have $\dnu{03} = \dnu{01}+\dnu{13}$, and we note that \dnu{03}
is the amount by which the $l=3$ mode is offset from the midpoint of the
adjacent $l=0$ modes.  This is arguably a more sensible measurement of the
position of the $l=3$ ridge than \dnu{13} because it is made with reference
to the radial modes \citep[see][]{kepler-bedding++2010-rg}.

Equation~\ref{eq.asymptotic} describes the oscillation frequencies from an
observational perspective.  A theoretical asymptotic expression
\citep{Tas80,Gou86,Gou2003} gives physical significance to \Dnu, \dnu{0l}
and $\epsilon$ as integrals of the sound speed.  It turns out that the
large separation is approximately proportional to the square root of the
mean stellar density, as discussed further in Sec.~\ref{sec.scaling}.  The
small separations in main-sequence stars are sensitive to the gradient of
the sound speed near the core, and hence to the age of the star (see
Sec.~\ref{sec.c-d}).  The asymptotic analysis (see previous references)
indicates that \dnu{0l} is proportional to $l(l+1)$.  Finally, $\epsilon$
is sensitive to upper turning point of the modes
\citep[e.g.,][]{Gou86,PH+ChD98,roxburgh2010-review}.  In the \'echelle
diagram, $\epsilon$ can be measured from the horizontal position of the
$l=0$ ridge.  In practice there is an ambiguity in~$\epsilon$ if $n$ is not
known (see Eq.~\ref{eq.asymptotic}).  In the Sun we know from theoretical
models that $\epsilon \approx 1.5$ (rather than 0.5), which allows us to
find the correct value of $n$ for each mode.  But it is important to
remember that $n$ is not directly observable.


\subsection{Departures from the asymptotic relation}
\label{sec.not.asymptotic}

The asymptotic relation in Eq.~\ref{eq.asymptotic} is only an
approximation.  Looking at Fig.~\ref{fig.bison.echelle}, we see that the
ridges all curve in an `S'~shape that is particularly pronounced at low
frequencies.  We also see that the $l=0$ and $l=2$ ridges converge towards
to top, as do the $l=1$ and $l=3$ ridges.  How should we best describe
these departures?  One approach is to say that \Dnu\ is a function of both
$l$ and frequency.  In fact, doing this is sufficient to describe all of
the features just mentioned.  However, it is physically more sensible to
say instead that the convergence of the ridges is due to changes of the
various {\em small\/} separations with frequency.  Indeed, a linear
dependence of $\dnu{02}$ on frequency fits the solar data quite well
\citep{helio-elsworth++1990-nature-core}, and has been adopted for other
stars \citep{white++2011-diagrams}.  If we allow the small separations to
vary with frequency then there is no longer any need to allow \Dnu\ to be a
function of~$l$.

What about the overall S-shaped curvature?  This can be described as a
change of \Dnu\ with frequency, and this is often done.  However, it seems
physically more sensible to ascribe the curvature to variations with
frequency of~$\epsilon$
\citep[e.g.,][]{Gou86,PH+ChD98,roxburgh2010-review}.  This allows \Dnu\ to
have a single value that is independent of both $l$ and frequency.
However, in practice if we are given a set of frequencies (either observed
or calculated) then \Dnu\ and $\epsilon$ must be measured together,
preferably in the region centred on \numax.  As we see from the open
symbols in Fig.~\ref{fig.bison.echelle}, a small change in \Dnu\ translates
to quite a large change in $\epsilon$.  These issues are discussed in some
detail by \citet{white++2011-diagrams}.



\section{Ground-based observations} \label{sec.ground}

\begin{figure}
\centerline{\includegraphics[width=\textwidth]{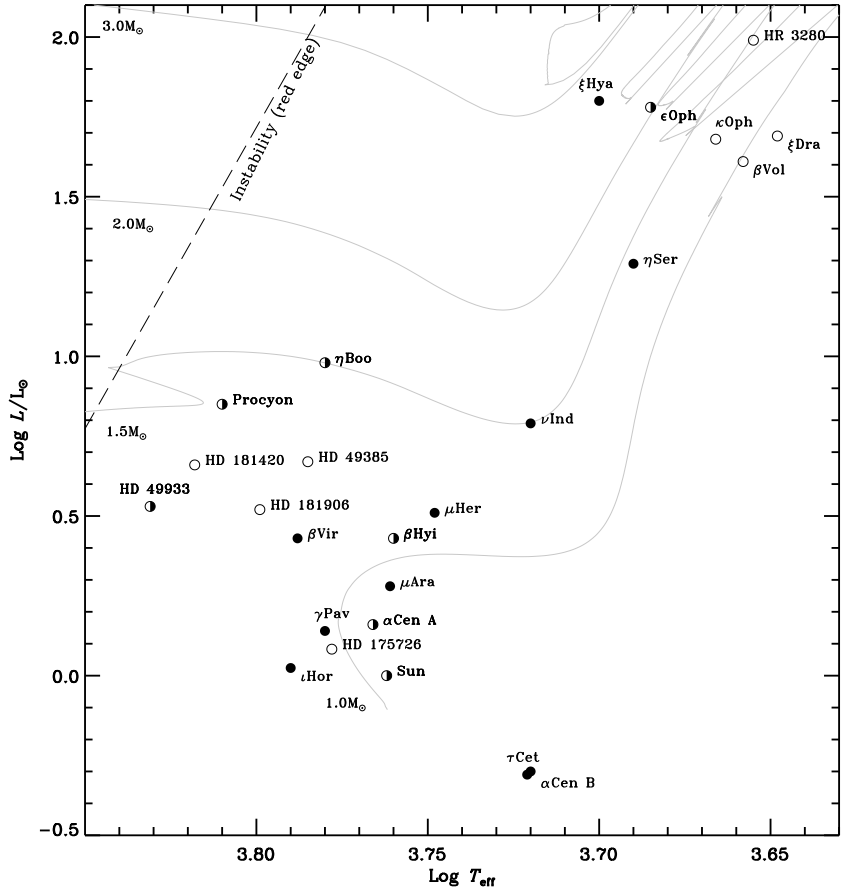}}
\caption{\label{fig.dennis} Hertzsprung-Russell diagram showing stars with
  solar-like oscillations for which \Dnu\ has been measured.  Filled
  symbols indicate observations from ground-based spectroscopy and open
  symbols indicate space-based photometry (some stars were observed using
  both methods).  Note that stars observed by \kepler\ are not included.
  Figure courtesy of Dennis Stello.  }
\end{figure}

We now review some of the observations of solar-like oscillations.  More
details and additional references can be found in various review articles
\citep[e.g.,][]{B+G94,HJLaB96,B+K2003,ChD2004,CAChD2007,BMS2008,B+K2008,AChDC2008}.
Figure~\ref{fig.dennis} is an HR diagram showing most of the stars
mentioned below.

\begin{figure}
\centerline{\includegraphics[width=0.7\textwidth]{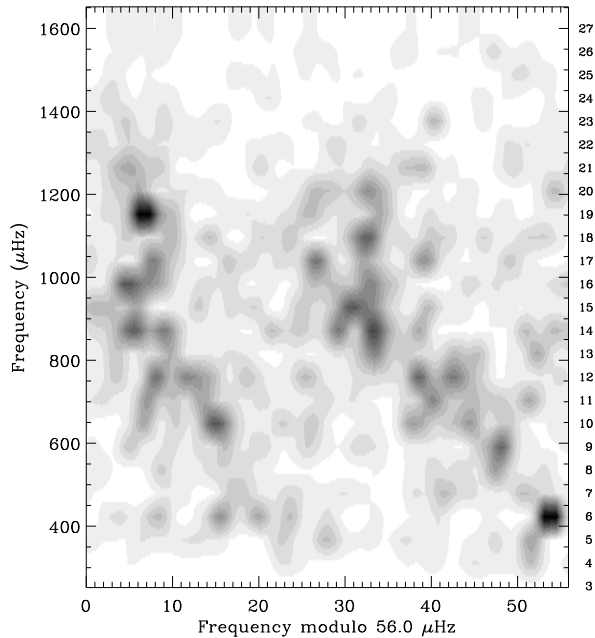}}
\caption{\label{fig.procyon.echelle} \'Echelle diagram of Procyon from the
multi-site campaign, showing broadening due to short mode lifetimes
\citep{BKC2010}. Which ridge is $l=1$ and which is $l=0$ and~2?  Note that
the power spectrum has been smoothed to a resolution of 2\,\muHz.}
\end{figure}

When global oscillations were discovered in the Sun, it was quickly
realised that measuring similar oscillations in Sun-like stars would be
tremendously valuable.  Several unsuccessful attempts were made
\citep[e.g.,][]{TMC78,BVH81,Smi82a} and some detections were claimed over
the next decade but a review by \citet{K+B95} concluded that none were
convincing.  However, it now seems clear that the excess power in the F5
star Procyon reported by \citet{BGN91} was indeed due to oscillations and
should be recognised as the first such detection, although it took some
time and more observations for the large separation (about 55\,\muHz) to be
established \citep{MMM98,MSL99,BMM99}.  Furthermore, even a multi-site
campaign on Procyon using eleven telescopes at eight observatories
\citep{AKB2008} has failed to give an unambiguous mode identification
\citep{BKC2010}, due to the short lifetimes of the oscillation modes (see
Fig.~\ref{fig.procyon.echelle}).  We refer to this as the F~star problem
(see also Sec.~\ref{sec.space}).

Most early oscillation searches measured Doppler shifts in radial
velocities.  However, a major effort to measure oscillations in intensity
was made in a multi-site photometric campaign on the open cluster M67 by
\citet{GBK93}.  Photometric measurements have the big advantage of allowing
simultaneous measurements of many stars This has now been done from space
by the \corot\ and \kepler\ missions, but ground-based photometry is
severely hampered by atmospheric scintillation.  The \citet{GBK93} M67
campaign did not produce definite detections, although it did indicate that
oscillation amplitudes of F~stars have lower amplitudes than expected from
theoretical calculations \citep{K+B95}.  It now seems clear that low
oscillation amplitudes in F~stars are a direct result of the short mode
lifetimes that were mentioned above for Procyon.


A third method for detecting oscillations was applied by \citet{KBV95} to
the G0 subgiant \eboo.  This method involved measuring fluctuations in the
equivalent widths of the temperature-sensitive Balmer lines using
low-resolution spectroscopy \citep[see also][]{BKR96}.  It produced good
evidence for multiple oscillation modes in \eboo\ with a large spacing of
40\,\muHz.  Subsequent observations confirmed this result
\citep{KBB2003,CEB2005}, establishing \eboo\ as the first star apart from
the Sun with a clear detection of solar-like oscillations.  A very
interesting result for \eboo, already apparent from the data obtained by
\citet{KBV95}, was the departure of some of the dipole ($l=1$) modes from
their regular spacing.  This was interpreted by \citet{ChDBK95} as arising
from bumping of mixed modes, as discussed in more detail in
Sec.~\ref{sec.mixed}.  Meanwhile, the equivalent-width method for observing
oscillations turned out to be more sensitive to instrumental drifts than we
hoped, although observations of \acena\ did produce tentative evidence for
p~modes \citep{KBF99}.

\begin{figure}
\includegraphics[width=0.8\textwidth]{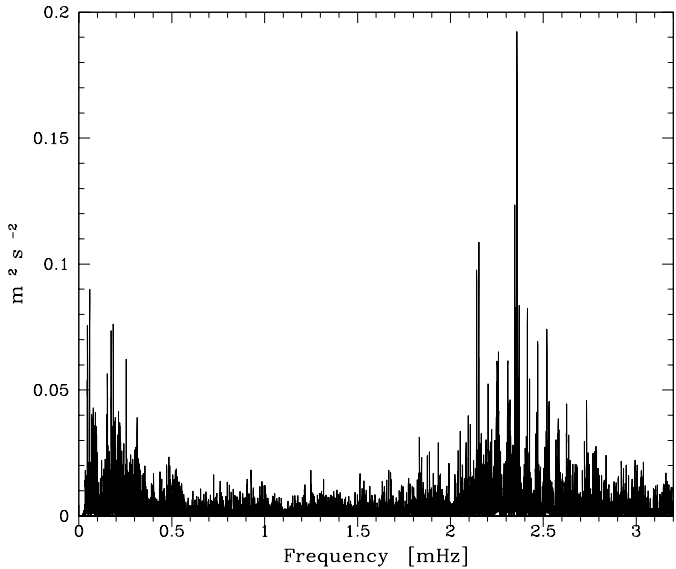}
\caption[]{\label{fig.acena} Power spectrum of \acena\ from velocity
  observations over 13 night with the CORALIE spectrograph by
  \citet{B+C2002}.}
\end{figure}

Major breakthroughs came from improvements in Doppler precision that were
driven by the push to find extra-solar planets.  Observations of the G2
subgiant star \bhyi, made with UCLES on the 3.9-m Anglo-Australian
Telescope \citep{BBK2001} and confirmed with CORALIE on the 1.2-m Swiss
Telescope \citep{CBK2001}, showed a clear power excess with a regular
spacing ($\Dnu = 56\,\muHz$).  This result was hailed by \citet{Gou2001} as
the birth of asteroseismology, but that honour should probably go to the
clear detection using CORALIE of the p-mode oscillation spectrum in \acena\
by \citet{B+C2001,B+C2002} (Fig.~\ref{fig.acena}).

Following on from these successes, a series of observations were made using
high-resolution spectrographs on medium-to-large telescopes.  Some
spectrographs were equipped with an iodine cell, such as UCLES at the AAT
(see above), UVES on the 8.2-m Very Large Telescope
\citep[e.g.,][]{BBK2004} and SARG on the 3.6-m Telescopio Nazionale Galileo
\citep[e.g.,][]{BBC2008}.  Other spectrographs used a thorium-argon lamp,
including CORALIE (see above), as well as HARPS on the European Southern
Observatory's 3.6-m telescope.  HARPS has produced some particularly
impressive results, including the first application of asteroseismology to
a planet-hosting star (\muara; see \citealt{BBS2005,BVB2005}), an age for
\ihor\ that implies it is an evaporated member of the primordial Hyades
cluster \citep{VLB2008}, and the detection of oscillations in the
remarkable solar twin 18~Sco \citep{BIH2011}.


Another important development was the clear detection of oscillations in
the red giant star \xihya\ based on one month of observations with CORALIE
\citep{FCA2002}.  This confirmed that red giants do indeed show solar-like
oscillations, as already suspected from earlier observations of Arcturus
and other stars.  However, the oscillation periods of several hours mean
that extremely long time series are needed, and the development of
red-giant asteroseismology had to wait until dedicated space missions, as
discussed in Sec.~\ref{sec.space}.

Single-site observations suffer from aliases, which complicate the
interpretation of the frequency spectrum (see Sec.~\ref{sec.fourier}).
Some stars have been observed from two sites, including \zher\
\citep{BSL2003}, Procyon \citep{MLA2004,KAS2008}, \acena\ \citep{BKB2004},
\acenb\ \citep{KBB2005}, \nuind\ \citep[a metal-poor
subgiant;][]{BBC2006,CKB2007}, and \bhyi\ \citep{BKA2007}.  The star \bhyi\
is a subgiant that, like \eboo, shows mixed $l=1$ modes that are
irregularly spaced (see Sec.~\ref{sec.mixed}).

Organising multi-site spectroscopic campaigns is difficult and the only
example so far was carried out on Procyon (see above).  The next step is
SONG, a dedicated network of 1-m telescopes equipped with iodine-stabilized
spectrographs \cite{GKF2006}.  The first node is currently under
construction on Tenerife.  Even in the era of space asteroseismology,
ground-based velocity measurements retain some advantages.  The background
noise from stellar granulation is lower than for intensity observations,
allowing better signal-to-noise at low frequencies.  There is also the
ability to cover the whole sky and to target nearby stars whose parallaxes
and other parameters are accurately known, such as the stars mentioned in
this section.


\section{Space-based observations} \label{sec.space}

Making photometric measurements from space avoids the scintillation caused
by the Earth's atmosphere and allows very high precision, even with a small
telescope.  Depending on the spacecraft orbit, it is possible to obtain
long time series with few interruptions.  The first space photometer
dedicated to asteroseismology was the {\em EVRIS\/} experiment aboard the
Russian {\em Mars~96\/} spacecraft.  {\em EVRIS\/} was intended to carry
out asteroseismology of bright stars during the cruise phase of the mission
\citep{Bag91,BAV97} but unfortunately the launch failed and the mission was
lost.

Other asteroseismology missions were proposed, including {\em PRISMA\/}
\citep{AGC93}, {\em STARS\/} \citep{FGJ95}, {\em MONS\/} \citep{KBChD2000}
and {\em Eddington\/} \citep{R+F2003}.  None of these made it though to
construction and the first success came from an unexpected direction.
After the failure of NASA's {\em Wide-Field Infrared Explorer\/} satellite
(\WIRE) immediately after launch due to the loss of coolant, its 50-mm star
camera was used for asteroseismology by \citet{Buz2002}.  \WIRE\ was able
to detect solar-like oscillations in \acena\ \citep{S+B2001,FCE2006} and
also in several red giants \citep{BCL2000,RBB2003,SBP2008}.  The SMEI
experiment (Solar Mass Ejection Imager) on board the {\em Coriolis\/}
satellite has also been used to measure oscillations in bright red giants
\citep{TCE2007,TCE2008}.  Finally, we note that the {\em Hubble Space
Telescope\/} has been used on several occasions to observe solar-like
oscillations, both via direct imaging \citep{E+G96,Gil2008,S+G2009} and
also using its Fine Guidance Sensors
\citep{ZKW99,rg-kallinger++2005-HST-FGS,gilliland++2011-HST-HD17156}.

The first dedicated asteroseismology mission to be launched successfully
was \MOST\ ({\em Microvariability and Oscillations of Stars}), which is a
15-cm telescope in low-Earth orbit \citep{WMK2003}.  It has achieved great
success with observations of classical (heat-engine) pulsators.  In the
realm of solar-like oscillations, controversy was generated when \MOST\
failed to find evidence for oscillations in Procyon \citep{MKG2004}.
However, \citet{BKB2005} argued that the \MOST\ non-detection was
consistent with the ground-based data.  Using photometry with \WIRE,
\citet{BKB2005b} extracted parameters for the stellar granulation and found
evidence for an excess due to oscillations.  The discrepancy was finally
laid to rest when \citet{HBA2011} found an oscillation signal in a new and
more accurate set of Procyon \MOST\ data that agreed well with results from
the ground-based velocity campaign (see Sec.~\ref{sec.ground}).

The French-led \corot\ mission, launched in December 2006, is a 27-cm
telescope in low-Earth orbit.  It has contributed substantially to the list
of main-sequence and subgiant stars with solar-like oscillations
\citep[e.g.,][]{AMA2008,MBA2008,BDB2009,GRS2009,corot-mosser++2009-hd175726,DBM2010,MGC2010}.
In particular, it has highlighted what might be called `the F~star problem'
in stars such as HD\,49933\
\citep[e.g.,][]{AMA2008,corot-benomar++2009-hd49933}, which is also a
problem in Procyon (\citealt{BKC2010}; see also \citealt{B+K2010}).  This
refers to stars in which the mode lifetimes are so short that the $l=0$ and
$l=2$ modes are blended, making them indistinguishable from $l=1$ modes.
This makes it difficult to decide which ridge in the \'echelle diagram
belongs to $l=1$ and which belongs to $l=0$ and~2, although measurement of
$\epsilon$ may be able to resolve this ambiguity (see
Sec.~\ref{sec.epsilon}).

A major achievement by \corot\ came from observations of hundreds G- and
K-type red giants that showed clear oscillation spectra that were
remarkably solar-like, with both radial and non-radial modes
\citep{DeRBB2009,HKB2009,CDeRB2010}.  \corot\ continues to produce
excellent results on red giants
\citep[e.g.,][]{MMB2009,KWB2010,MBG2010,MBG2011}, and also on main-sequence
and subgiant stars
\citep[e.g.,][]{corot-mathur++2010-hd170987,corot-ballot++2011-hd52265}

Finally we come to \kepler, which began science observations in May 2009.
It has a 95-cm aperture and operates in an Earth-trailing heliocentric
orbit, allowing continuous observations without interference from reflected
light from the Earth.  Although its primary goal is to search for
extra-solar planets \citep{kepler-borucki++2010-science}, it is no
exaggeration to say that \kepler\ is revolutionising asteroseismology
\citep{kepler-gilliland++2010-pasp}.  Most stars are observed in
long-cadence mode (sampling 29.4 minutes; see
\citealt{jenkins++2010-long-cadence}), which is adequate to study
oscillations in most red giants but is not fast enough for Sun-like stars.
Fortunately, about 500 stars at a time can be observed in short-cadence
mode (sampling 58.8\,s; see \citealt{kepler-gilliland++2010-short-cadence}).

For the first seven months of science operations, a survey was carried out
with {\em Kepler's\/} short-cadence mode that targeted more than 2000
main-sequence and subgiant stars for one month each.  The survey detected
solar-like oscillations, including a clear measurement of \Dnu, for about
500 stars \citep{kepler-chaplin++2011-science}, representing an increase by
a factor of $\sim$20 over the situation prior to \kepler.

Some of these stars have been studied individually.  The first three to be
published \citep{kepler-chaplin++2010-solar-like3} were `Java'
(KIC~3656476), `Saxo' (KIC~6603624) and `Gemma'
(KIC~11026764).\footnote{Within the \kepler\ working groups, some stars are
known by nicknames for convenience.}  Java and Saxo are main-sequence
stars, while Gemma is a subgiant showing bumped $l=1$ modes
(\citealt{kepler-metcalfe++2010-gemma}; see Sec.~\ref{sec.mixed} for a
discussion of mode bumping).  A small fraction of the survey stars were
observed continuously since the start of the mission\footnote{Note that one
of the 22 CCD modules on \kepler\ failed after seven months of operation.
Since the spacecraft rotates by 90 degrees every three months to maintain
the orientation of the solar panels, about 20\% of stars will fall on this
failed module at some time during the year, creating annual 3-month gaps in
the data for those stars.}.  Of these, results have so far been published
on `Tigger' (KIC~11234888) and `Boogie' (KIC~11395018) by
\citet{kepler-mathur++2011-boogie-tigger}, and on `Mulder' (KIC~10273246)
`Scully' (KIC~10920273) by \citet{kepler-campante++2011-mulder-scully}.


As mentioned, the long-cadence mode of \kepler\ is perfectly adequate for
red giants, and solar-like oscillations have been detected in thousands of
them
\citep[e.g.,][]{kepler-bedding++2010-rg,kepler-huber++2010-rg,kepler-kallinger++2010-rg,kepler-hekker++2011-rg-public},
including some in open clusters
\citep{kepler-stello++2010,kepler-basu++2011,kepler-hekker++2011-clusters}.
Some of these results are discussed in more detail in
Sec.~\ref{sec.redgiants} (see also reviews by
\citealt{rg-hekker2010-review} and \citealt{ChD2011}).


\section{A few comments about Fourier analysis} \label{sec.fourier}

Fourier analysis is an indispensable tool for asteroseismology.  This
section introduces some of the most important aspects.  For more details
including full references see, for example, \citet{Pijpers2006},
\citet{AChDK2010} and \citet{App2011}.

The {\em amplitude spectrum\/} of a time series tells us which frequencies
are present, and with which amplitudes.  Note that the {\em power
spectrum\/} is simply the square of the amplitude spectrum.  A useful way
of thinking about the amplitude spectrum is as follows.  We choose a
frequency~$\nu$ and perform a least-squares fit of a sine wave to the time
series.  That is, we fix the frequency of the sine wave to be $\nu$ and we
vary the amplitude and phase until we get the best fit.  If the quality of
the time series is not uniform, as always happens with ground-based
observations, it is important to weight the data using the measurement
uncertainties \citep[e.g.,][]{FJK95}.  The amplitude of this best-fitting
sine wave is our result (for solar-like oscillations, which are
stochastically excited, we are usually not interested in the phase).
Repeating this calculation for a range of different frequencies generates
the amplitude spectrum.

This way of describing the amplitude spectrum allows us to understand
several important properties.  Firstly, it is clear that the observations
do not need to be equally spaced in time (and ground-based observations
generally are not).  Now suppose our time series consists of a single
sinusoidal signal.  If we choose the correct frequency then the
best-fitting amplitude will match the amplitude of the signal.  If we
choose completely the wrong frequency, the best-fitting sine wave will have
a very small amplitude.  But if we choose a frequency that is only slightly
different from that of the signal, the best-fitting amplitude will be
slightly less than the signal amplitude.  Repeating this in small frequency
steps will trace a sinc function ($\sin\nu/\nu$).  Increasing the length of
the time series will decrease the width of this sinc function.

If there are regular gaps in the time series, as occurs for ground-based
single-site observations, then choosing the wrong frequency will still give
a significant response in some cases.  For example, suppose the sine wave
we are fitting has a higher frequency than the signal, such that exactly
one extra cycle fits into the gap.  This sine wave will still give a fairly
good fit to the data and we will get an extra peak in the amplitude
spectrum, called a {\em sidelobe}.  The same thing occurs if the chosen
frequency is exactly one cycle-per-gap {\em lower\/} than the signal
frequency.  We therefore expect two sidelobes, one on either side of the
true peak.  Sidelobes are an example of {\em aliasing}, which occurs when
two or more different sine waves give good fits to the same data.  The
heights of the sidelobes relative to the true peak will decrease as the
gaps get smaller (and the duty cycle improves).  If the period of the gaps
in the time series is $T_{\rm gap}$ then the sidelobes of a signal having
frequency $\nu$ will occur at frequencies of $\nu \pm 1/T_{\rm gap}$.
There will also be a smaller pair of peaks separated from the true peak by
$\nu \pm 2/T_{\rm gap}$, another pair at $\nu \pm 3/T_{\rm gap}$, and so
on.

\begin{figure}
\centerline{\includegraphics[width=0.9\textwidth]{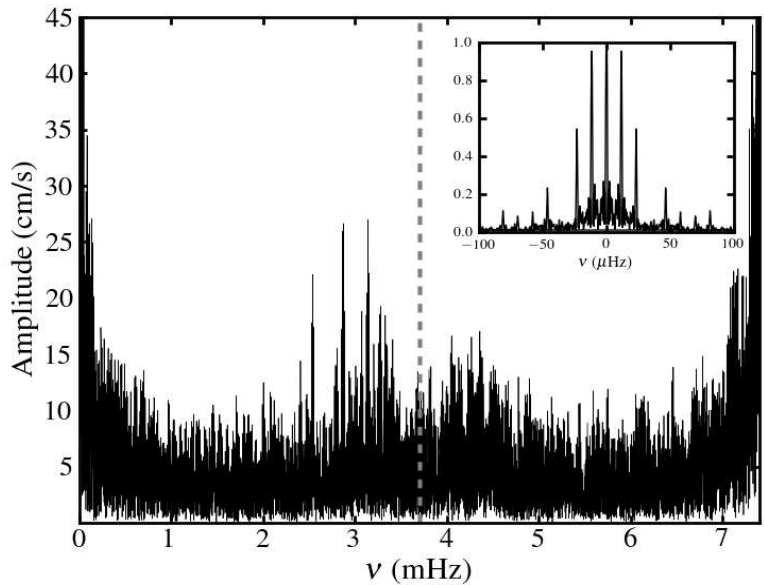}}
\caption{\label{fig.18sco} Amplitude spectrum of the solar twin 18~Sco,
  based on observations taken with HARPS over 12 nights.  The inset shows
  the spectral window in amplitude, and the vertical dashed line shows the
  effective Nyquist frequency.  Figure from \citet{BIH2011}.  }
\end{figure}

This leads to the concept of the {\em spectral window}, which is the power
spectrum of a pure sine wave.  It is sometimes incorrectly called the {\em
window function}, but that term actually refers to the sampling in the time
domain.  In practice, the spectral window should be calculated by averaging
the power spectra of a sine and a cosine wave, to eliminate end effects.
Note that this averaging should be done in power and not in amplitude.  The
inset of Fig.~\ref{fig.18sco} shows an example.

The another example of aliasing occurs if the observations are regularly
sampled.  A single sine wave with frequency $\nu$ that is sampled regularly
at intervals separated by $\delta t$ can be fitted equally well by a much
higher-frequency sine wave having frequency $2 \nu_{\rm Nyq} - \nu$, where
$\nu_{\rm Nyq} = 1/(2\delta t)$ is called the Nyquist frequency.  This
aliasing still occurs, although with reduced amplitude, if the sampling is
not exactly regular.  An example is shown in Fig.~\ref{fig.18sco} for
ground-based data.  In this case, one can define an effective Nyquist
frequency based on the median sampling.


It is sometimes assumed that the power spectrum does not contain any useful
information above the Nyquist frequency, and that any signal whose
frequency is higher than $\nu_{\rm Nyq}$ cannot be measured.  This is not
true.  For example, observations taken with \kepler\ in its long-cadence
mode (sampling 29.4 minutes, $\nu_{\rm Nyq} = 283\,\muHz$) still contain
useful information about oscillations at higher frequencies, provided the
power spectrum is interpreted with care.
 



\section{Scaling relations for \Dnu\ and \numax} \label{sec.scaling}

As mentioned in Sec.~\ref{sec.asymptotic}, the large separation scales
approximately with the square root of the mean stellar density:
\begin{equation}
 \frac{\Dnu}{\Dnu_\odot} \approx \sqrt{\frac{\rho}{\rho_\odot}} =
  \left(\frac{M}{M_\odot}\right)^{0.5}
 \left(\frac{\Teff}{\Teff_\odot}\right)^{3} \left(\frac{L}{L_\odot}\right)^{-0.75}. \label{eq.Dnu.scaling}
\end{equation}
This scaling relation is very widely used, but how accurate is it?  It was
recently tested using theoretical models by \citet{white++2011-diagrams}.
Of course, given the curvature in the \'echelle diagram, the measurement of
\Dnu\ depends on the frequency at which it is measured (see
Sec.~\ref{sec.asymptotic}).  The approach adopted by
\citet{white++2011-diagrams} was to fit to the radial modes near \numax\
(see the paper for details).  They confirmed generally good agreement with
Eq.~\ref{eq.Dnu.scaling} but found departures reaching several percent.
Interestingly, they found that these departures are predominantly a
function of effective temperature and suggested a revised scaling relation
that takes this into account.

Even if Equation~\ref{eq.Dnu.scaling} (or a modified version) agrees with
theoretical models, this does not necessarily mean it is accurate for real
stars.  There are well-known discrepancies between the observed and
calculated oscillation frequencies in the Sun, which are due to incorrect
modeling of the surface layers \citep{ChDDL88,DPV88,RChDN99,LRD2002}.
These discrepancies increase with frequency so there must also be a
discrepancy in~\Dnu.  Indeed, the value of \Dnu\ of the best-fitting solar
model is about 1\% greater than the observed value \citep{KBChD2008}.  We
can hardly expect the situation to be better in other stars, and a solution
to the problem of modelling near-surface layers is urgently needed.

\begin{figure}


\centerline{
\includegraphics[width=0.7\textwidth]{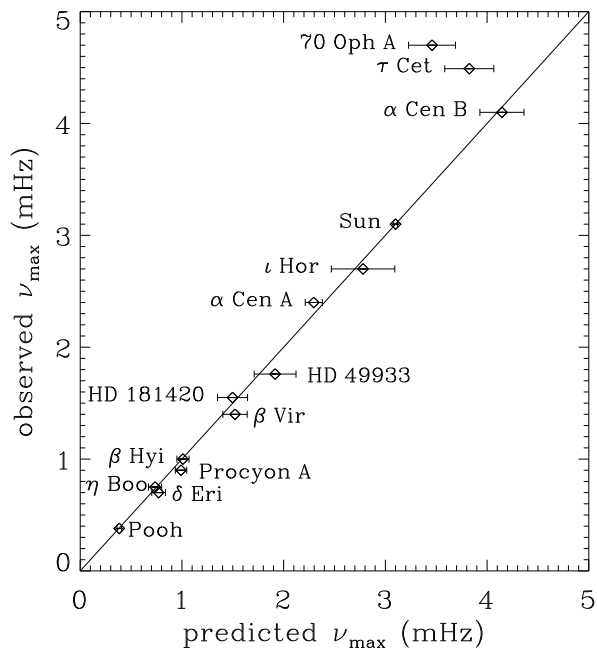}} 

\caption{\label{fig.numax} Testing the scaling relation for \numax.  Values
for the stellar parameters were taken from the compilations by
\citet{corot-bruntt2009} and \citet{bruntt++2010-parameters}, with the
addition of `Pooh' (KIC~4351319), a low-luminosity red giant observed by
\kepler\ \citep{kepler-dimauro++2011-pooh}. }
\end{figure}

Another important and widely used scaling relation expresses \numax\ in
terms of stellar parameters.  It is based on the suggestion by
\citet{BGN91} that \numax\ might be expected to be a fixed fraction of the
acoustic cutoff frequency, which in turn scales as $\Teff/\sqrt{g}$.  Hence
we have:
\begin{equation}
 \frac{\numax}{\numax_{,\odot}} \approx \frac{M}{M_\odot}
 \left(\frac{\Teff}{\Teff_\odot}\right)^{3.5} \left(\frac{L}{L_\odot}\right)^{-1}. \label{eq.numax.scaling}
\end{equation}
This relation has been shown to agree quite well with observations
\citep{B+K2003} and also with model calculations \citep{CHA2008}, and has
recently been given some theoretical justification by \citet{BGD2011}.
Figure~\ref{fig.numax} shows the expected and observed values of \numax\
for some well-studied stars with accurate parallaxes.  There is generally
very good agreement except for the two low-mass stars \taucet\ and
70~Oph~A.  If confirmed, this may indicate a breakdown of the assumption
that \numax\ is a fixed fraction of the acoustic cutoff frequency.

\section{Ensemble asteroseismology}

The wealth of data from \corot\ and \kepler\ allow us to carry out what
might be called {\em ensemble asteroseismology}.  This involves studying
similarities and differences in groups of stars, and is complementary to
the detailed study of a few individuals.  Several groups have developed
pipelines to process the large amounts of data and extract the global
parameters such as \numax, \Dnu\ and amplitude.  These various pipelines
have been tested and compared using both simulations
\citep{chaplin++2008-asteroflag1} and \kepler\ data
\citep{kepler-hekker++2011-rg-comparison,kepler-verner++2011-comparison}.


Ensemble asteroseismology can be carried out using asteroseismic diagrams,
in which two properties of the oscillation spectra are plotted against one
another.  Here, we briefly discuss some of these diagrams.

\paragraph{The C-D diagram:}
\label{sec.c-d}

\begin{figure}
\centerline{\includegraphics[width=1.0\textwidth]{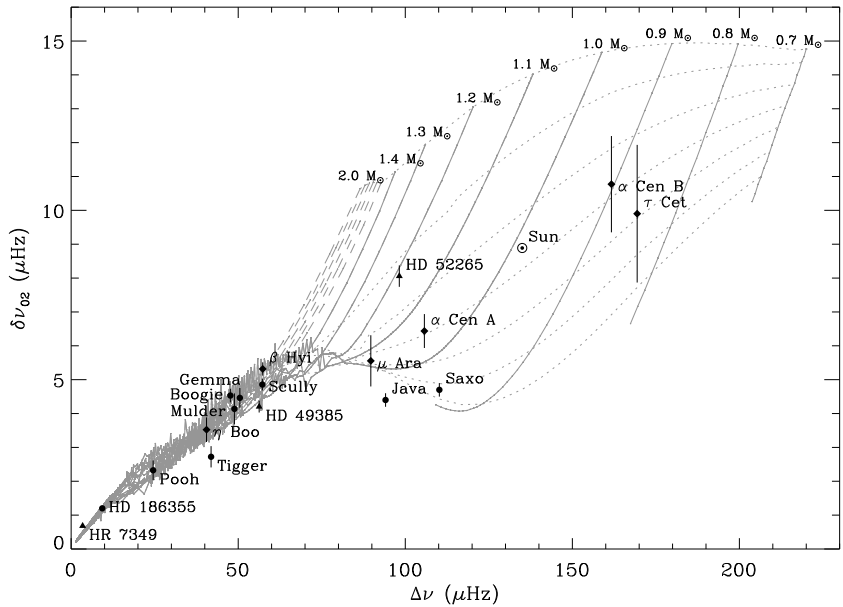}}
\caption{\label{fig.CD} The so-called C-D diagram, which plots the large
  and small frequency separations.  The solid lines show model calculations
  for solar metallicity using ASTEC and ADIPLS \citep{ChD2008a,ChD2008b},
  with the dashed lines being isochrones from zero to 12\,Gyr (top to
  bottom) in steps of 2\,Gyr.  The points show measurements of stars
  showing solar-like oscillations.  Figure adapted from
  \citet{white++2011-diagrams}.  }
\end{figure}

The plot of small versus large frequency separations, first introduced by
\citet{ChD84}, has come to be known as the C-D diagram.  A modern version
is shown in Fig.~\ref{fig.CD}.  In main-sequence stars, the tracks for
different masses and ages are well spread, allowing both to be read off the
diagram.  However, note that the positions of the tracks depend on
metallicity, and so this must be taken into account \citep[for details and
references, see][]{white++2011-diagrams}.  For more evolved stars, the
tracks converge and the small separation becomes much less sensitive as a
diagnostic.  C-D diagrams for red giants observed with \kepler\ have been
presented by \citet{kepler-bedding++2010-rg} and
\citet{kepler-huber++2010-rg}.

\paragraph{The \numax-\Dnu\ diagram:}

\begin{figure}
\centerline{\includegraphics[width=0.8\textwidth]{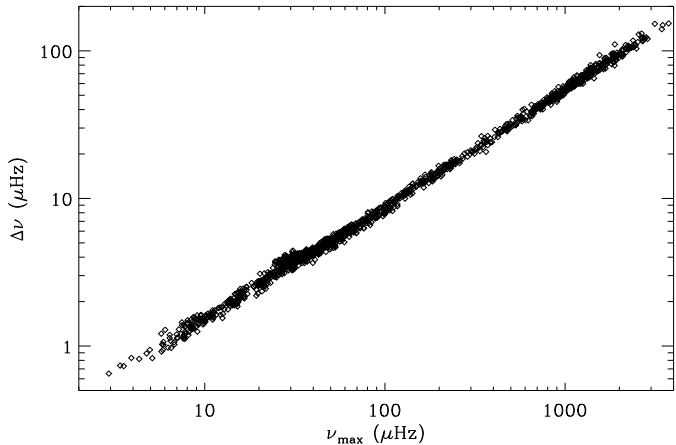}}
\caption{\label{fig.numax-dnu} The strong correlation between \numax\ and
  \Dnu, based on \kepler\ observations from the main sequence (upper right)
  to the red giants (lower left).  Figure adapted from
  \citet{kepler-huber++2011-ensemble}.  }
\end{figure}

The strong correlation between \numax\ and \Dnu\ was first plotted for
main-sequence and subgiant stars by \citet{SCB2009} and for red giants by
\citet{HKB2009}.  The tightness of this relationship follows from the
scaling relations (Eqs.~\ref{eq.Dnu.scaling} and~\ref{eq.numax.scaling}),
as discussed in some detail by \citet{SCB2009}.  The relationship extends
over several orders of magnitude (see Fig.~\ref{fig.numax-dnu}), making it
a useful predictor of~\Dnu.

\paragraph{The $\epsilon$ diagram:} \label{sec.epsilon}

As discussed in Sec.~\ref{sec.asymptotic}, the parameter $\epsilon$ is
sensitive to the surface layers of the star but needs to be measured
carefully.  A diagram showing $\epsilon$ versus \Dnu\ was calculated by
\citet{ChD84}, but its diagnostic potential was not realised until
recently.  \citet{B+K2010} suggested that $\epsilon$ could be used as an
aid to mode identification (the `F~star problem'; see
Secs.~\ref{sec.ground} and ~\ref{sec.space}).  The first observationl
$\epsilon$ diagram was made for red giants observed with \kepler\ by
\citet{kepler-huber++2010-rg}, who used it to suggest mode identifications
for four \corot\ red giants discussed by \citet{corot-hekker++2010-rg}.
Meanwhile, \citet{white++2011-diagrams} have shown that $\epsilon$ is
mostly determined by the stellar effective temperature and demonstrated
that it is indeed useful for addressing the F~star problem.


\section{Mode bumping and avoided crossings}
\label{sec.mixed}

As mentioned in Sec.~\ref{sec.modes}, mixed modes have p-mode character in
the envelope and g-mode character in the core.  They occur in evolved stars
(subgiants and red giants), in which the large density gradient outside the
core effectively divides the star into two coupled cavities.  This leads to
the phenomenon of {\em mode bumping}, in which mode frequencies are shifted
from their regular spacing.  Here, we summarise the main features of mixed
modes from an observational perspective, which necessarily includes a basic
introduction to the theory.  For more details on the theoretical aspects
see, for example,
\citet{Scu74,Osa75,ASW77,D+P91,ChD2004,MME2008,DBS2009,AChDK2010}.

\begin{figure}

\centerline{
\includegraphics[angle=90,height=0.25\textheight]{figs/fig.F3-un-tim.epsi}
\includegraphics[angle=90,height=0.25\textheight]{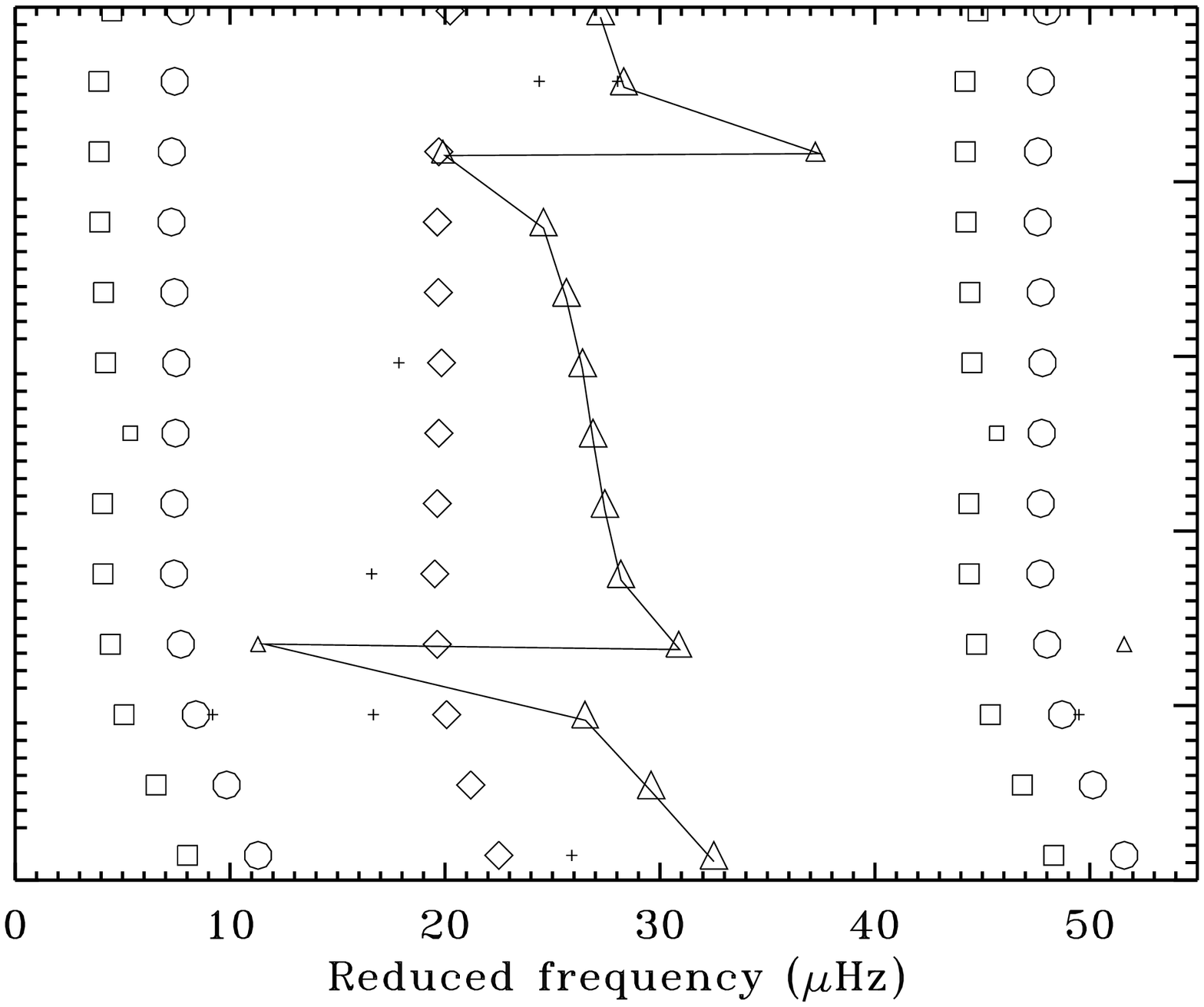}}

\caption{\label{fig.etaboo} Left: Evolution of oscillation frequencies in
models of a subgiant star of mass $1.60 \Msun$ and $Z = 0.03$ (representing
\eboo).  The dashed lines correspond to modes of degree $l = 0$, and the
solid lines to $l = 1$.  The vertical solid line indicates the location of
the model whose frequencies are illustrated in the right panel, the thick
solid lines show two of the $l=1$ $\pi$~modes and the thick dashed line
shows one of the $\gamma$~modes (see text).  Right: \'Echelle diagram using
a frequency separation of $\Delta\nu = 40.3 \muHz$ and a zero-point
frequency of $\nu_0 = 735 \muHz$.  Circles are used for modes with $l = 0$,
triangles for $l = 1$, squares for $l = 2$ and diamonds for $l = 3$.  The
size of the symbols indicates the relative amplitude of the modes,
estimated from the mode inertia, with crosses used for modes whose symbols
would otherwise be too small.  Figures adapted from \citet{ChDBK95}. }
\end{figure}

Figure~\ref{fig.etaboo} shows theoretical oscillations frequencies for a
subgiant star whose parameters were chosen to match \eboo.  The left panel
shows the evolution with time of the model frequencies for modes with $l=0$
(dashed lines) and $l=1$ (solid lines).  The $l=0$ frequencies decrease
slowly with time as the star expands.  At a given moment in time, such as
that marked by the vertical line, the radial modes are regularly spaced in
frequency, with a large separation of $\Dnu \approx 40\,\muHz$.  However,
the behaviour of the $l=1$ modes (solid lines) is rather different.  They
undergo a series of {\em avoided crossings\/} \citep{Osa75,ASW77}, during
which an $l=1$ mode is bumped upwards by the mode below, and in turn it
bumps the mode above.  The effect is to disturb the regular spacing of the
$l=1$ modes in the vicinity of each avoided crossing.  The right panel of
Fig.~\ref{fig.etaboo} shows the frequencies in \'echelle format for the
single model that is marked in the left panel by the vertical line.  The
bumping from regularity of the $l=1$ modes (triangles) is obvious and there
is one extra $l=1$ mode at each avoided crossing, where the modes are
``squeezed together''.

Bumping of $l=1$ modes has been observed in the subgiant stars \eboo\
\citep{KBV95,KBB2003,CEB2005}, \bhyi\ \citep{BKA2007}, the \corot\ target
HD~49385 \citep{DBM2010} and the \kepler\ target `Gemma' (KIC~11026764;
\citealt{kepler-metcalfe++2010-gemma}).  Many other subgiants stars
observed with \kepler\ also show bumping of $l=1$ modes, and observations
have so far been published for `Tigger' (KIC~11234888) and `Boogie'
(KIC~11395018) by \citet{kepler-mathur++2011-boogie-tigger}, and for
`Mulder' (KIC~10273246) and `Scully' (KIC~10920273) by
\citet{kepler-campante++2011-mulder-scully}.  It was also suggested by
\citet{BKC2010} that Procyon has an $l=1$ mixed mode at low frequency (see
Fig.~\ref{fig.procyon.echelle}), based on the narrowness of the peak in the
power spectrum.  Note that mixed modes are expected to have longer
lifetimes (smaller linewidths) than pure $p$ modes because they have larger
mode inertias \citep[e.g.,][]{ChD2004}.  Mode inertia is a measure of the
total interior mass that is affected by the oscillation.

We now discuss the theoretical aspects of mixed modes in more detail.  In
particular, what causes the mode bumping and why are $l=1$ modes the most
affected?  To answer these questions, we return to the idea of a star with
two cavities, as mentioned above.  We can think of subgiants and red giants
as having p~modes trapped in the envelope and g~modes trapped in the core.
Strictly speaking, these are hypothetical modes within each cavity rather
than true oscillation modes of the whole star, and so they should probably
be called `envelope p~modes' and `core g~modes'.  \citet{ASW77} referred to
these artificially isolated modes as $\pi$~modes and $\gamma$~modes, which
is a useful abbreviation that we will adopt.  They are the modes that would
exist if the two cavities were completely decoupled.  In reality, there is
coupling and this leads to a mixed mode whenever a $\pi$~mode and a
$\gamma$~mode with the same $l$ are sufficiently close to each other in
frequency.  This mixed mode has p-mode character in the envelope and g-mode
character in the core.  Note also that the frequencies of the $\pi$~modes
decrease with time as the star evolves, while those of the $\gamma$~modes
increase with time.

This picture allows to understand the features seen in
Fig.~\ref{fig.etaboo}.  For $l=0$ there are no mixed modes because radial
g~modes do not exist and so we have pure p~modes that are regularly spaced
in frequency.  For $l=1$, the $\gamma$~modes are visible as the three
upward-sloping ridges, one of which is marked with a thick dashed line
\citep[see also Fig.~1b of][]{ASW77}.  The $l=1$ $\pi$~modes can also be
traced in Fig.~\ref{fig.etaboo} by connecting the downward-sloping parts of
the tracks, as shown for two adjacent $\pi$~modes by the thick solid lines.
Recall that a mixed mode occurs when a $\pi$ mode and a $\gamma$ mode are
close to each other in frequency, which happens at the intersections.  In
fact, we see that several $\pi$ modes are able to couple with each $\gamma$
mode \citep[see][]{deheuvels+michel2010-crossings}.  Indeed, all the $l=1$
modes are bumped to some extent, with those closest in frequency to each
$\gamma$ mode being bumped the most, as seen in the \'echelle diagram
(right panel of Fig.~\ref{fig.etaboo}).\footnote{``All modes are mixed, but
some modes are more mixed than others'' \citep[with apologies to
George][]{Orwell-Animal-Farm}.}

\begin{figure}
\centerline{\includegraphics[width=0.8\textwidth]{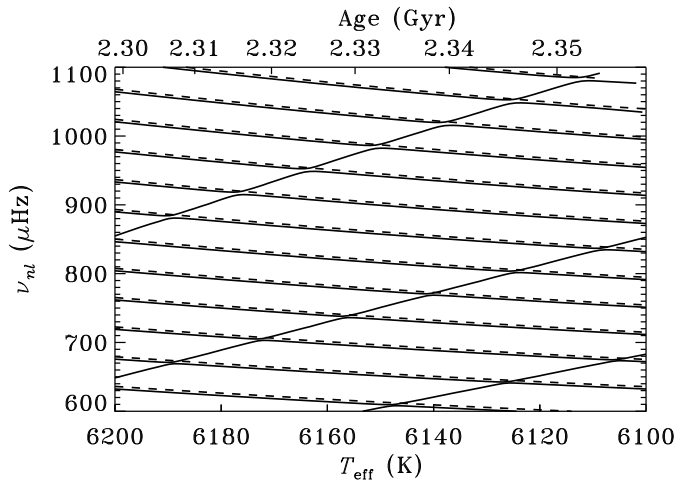}}

\caption{\label{fig.etaboo2} Avoided crossings for $l=2$ modes in a model
  of \eboo.  This is similar to the left panel of Fig.~\ref{fig.etaboo},
  but for a slightly different model and with the solid lines now showing
  $l=2$ modes (the dashed lines again show $l=0$ modes).  Adapted from
  \citet{ChD+H2010}.}
\end{figure}

Figure~\ref{fig.etaboo2} shows the situation for the $l=2$ modes in a
similar set of models.  The coupling between the two cavities for $l=2$
modes is much weaker than for $l=1$ modes,%
\footnote{The coupling strength varies inversely with the width of the
evanescent region between the two cavities, and this region turns out to be
wider for $l=2$ modes than for $l=1$ modes \citep{ChD2004}.}
which means that the $\pi$ and $\gamma$ modes must have almost exactly the
same frequency in order to couple.  In Fig.~\ref{fig.etaboo2} we see that
the $l=2$ modes (solid lines) mostly slope downward (following the
$\pi$-mode tracks), indicating they are essentially behaving as pure
p~modes.  The upward-sloping segments (which follow the $\gamma$-mode
tracks) correspond to modes behaving essentially as pure g~modes.  These
g-dominated modes will have very high inertias and very low amplitudes,
making them undetectable.  As a result, the observable $l=2$ modes will
only be bumped slightly or not at all, as can be seen in the right panel of
Fig.~\ref{fig.etaboo} (squares).

\subsection{The p-g diagram} \label{sec.pg-diagram}

It has long been recognised that mixed modes have great potential for
asteroseismology because their frequencies are very sensitive to stellar
interiors and because they change quite rapidly as the star evolves
\citep[e.g.,][]{D+P91,ChDBK95}.  One way to exploit this information is by
performing a direct comparison of observed frequencies with models.  This
has been done for \eboo\ \citep{ChDBK95,G+D96,DiMChDK2003,Gue2004,CEB2005},
\bhyi\ \citep{DiMChDP2003,F+M2003,brandao++2011-bhyi}, the \corot\ target
HD~49385 \citep{deheuvels+michel2010-crossings} and the \kepler\ target
`Gemma' \citep{kepler-metcalfe++2010-gemma}.

However, adjusting the model parameters to fit the observed frequencies can
be difficult and time-consuming.  We can ask whether the information
contained in the mixed modes can be used more elegantly.  Based on the
preceding discussion of mixed modes and mode bumping, we are led to
consider the frequencies of the avoided crossings themselves.
%
%
For example, the \'echelle diagram in Fig.~\ref{fig.etaboo} shows two
avoided crossings\footnote{Strictly speaking, these features in the
\'echelle diagram should not really be called avoided crossings, since that
term refers to the close approaches of two modes in the evolution diagram.
However, there does not seem to be any harm in using the term.}, at about
730 and 1050\,\muHz.  We recognise these as the frequencies of the $\gamma$
modes (recall that these are the pure g~modes that would exist in the core
cavity if it could be treated in isolation).  Much of the diagnostic
information contained in the mixed modes can be captured in this way.  This
is because the overall pattern of the mixed modes is determined by the mode
bumping at each avoided crossing, and these patterns are determined by the
g~modes trapped in the core (the $\gamma$~modes).  For related discussion
on this point, see \citet{deheuvels+michel2010-mu-gradient}.

This discussion suggests a new asteroseismic diagram, inspired by the
classical C-D diagram, in which the frequencies of the avoided crossings
(the $\gamma$ modes) are plotted against the large separation of the
p~modes.  This p-g diagram, so named because it plots g-mode frequencies
versus p-mode frequencies, could prove to be an instructive way to display
results of many stars and to make a first comparison with theoretical
models (\citealt{kepler-campante++2011-mulder-scully}; Bedding et al., in
prep.).


\section{Solar-like oscillations in red giants} \label{sec.redgiants}

As outlined in Sec.~\ref{sec.space}, there has recently been tremendous
progress in asteroseismology of G- and K-type red giants, with solar-like
oscillations detected by \corot\ and \kepler\ in thousands of stars.  There
have also been rapid developments in the theory
\citep[e.g.,][]{GDB2000,DGH2001,ChD2004,DBS2009,MMN2010,kepler-dimauro++2011-pooh},
as discussed in detail by \citet{ChD2011} in this volume.

At first glance, the oscillation spectra of red giants are remarkably
solar-like.  As first revealed by \corot\ observations \citep{DeRBB2009},
they have radial and non-radial modes that closely follow the asymptotic
relation.  In fact, to a large extent their power spectra are simply scaled
versions of each other, so that the large and small spacings are in a
roughly constant ratio \citep[e.g.,][]{kepler-bedding++2010-rg}.  This high
degree of homology is rather unhelpful for asteroseismology, in the sense
that the small separations lack much of the diagnostic potential that makes
them so valuable in main-sequence stars.  Fortunately, there are other
features in the oscillation spectra of red giants that make up for this
loss.  We now discuss one of them in some detail, namely the presence of
mixed modes with g-mode period spacings.

Even after the first month, \kepler\ data revealed multiple $l=1$ modes per
order that indicated the presence of mixed modes \citep[see Fig.~5
of][]{kepler-bedding++2010-rg}.  As with the subgiants discussed in
Sec.~\ref{sec.mixed}, the large density gradient outside the helium core
divides the star into two coupled cavities, with the oscillations behaving
like p~modes in the envelope and like g~modes in the core.  The difference
with red giants is that the spectrum of g~modes in the core (the
$\gamma$~modes) is much denser.  It is also important to note that, while
p~modes are approximately equally spaced in frequency (see
Sec.~\ref{sec.spectrum}), an asymptotic analysis shows that g~modes are
approximately equally spaced in {\em period\/} \citep{Tas80}.

\begin{figure}
\centerline{\includegraphics[width=\textwidth]{figs/evolution9-tim.epsi}}
\caption{\label{fig.rg}   }
\end{figure}

Figure~\ref{fig.rg}a shows the time evolution of frequencies of a red giant
model for modes with $l=0$ (dashed lines) and $l=1$ (solid lines).  It is
very instructive to compare this with Fig.~\ref{fig.etaboo}.  In both cases
we can follow the downward-sloping $l=1$ $\pi$~modes (two examples are
shown by thick solid lines).  They are approximately equally spaced in
frequency (with separation \Dnu) and they run parallel to the $l=0$ modes
(which are pure p~modes).  In both figures we can also follow the
upward-sloping $l=1$ $\gamma$~modes (examples shown by thick dashed lines).
The difference is that for the subgiant in Fig.~\ref{fig.etaboo}, there are
only three $\gamma$~modes, but for the red giant in Fig.~\ref{fig.rg}a
there are dozens.  As before, we have a mixed mode for every $\pi$~mode and
a mixed mode for every $\gamma$~mode.  Most of these mixed modes are
g-dominated (they follow the $\gamma$-mode tracks), with very high inertia
and very low amplitude.  However, some mixed modes fall close enough to a
$\pi$~mode for the coupling to be significant, producing p-dominated mixed
modes with sufficient amplitude to be observable.  These are shown by the
triangles in Fig~\ref{fig.rg}b, which is the \'echelle diagram of the model
indicated by the vertical line in Fig.~\ref{fig.rg}a.  Note that the
clusters of observed $l=1$ modes are centred at the positions expected for
pure $l=1$ p~modes, which are about half way between the $l=0$ modes.

The pattern of $l=1$ clusters described above has been observed in red
giants using data from \kepler\
\citep{kepler-beck++2011-g-modes,kepler-bedding++2011-rg-nature} and
\corot\ \citep{corot-mosser++2011-mixed}.  The spacings within the $l=1$
clusters clearly provide valuable information about the gravity modes
trapped in the core but to interpret them we must take into account the
mode bumping.  Each cluster is squeezed together by the presence of one
extra mode and so the spacing we measure from the observable modes, which
we denote \DPobs, will be substantially less than the spacing of the core
g~modes (the $\gamma$~modes), which we denote~\DPg.  This is shown for the
model in Fig.~\ref{fig.rg}c, which plots the period spacing between
consecutive $l=1$ mixed modes.  From the upper envelope we see that $\DPg
\approx 75$\,s, but this is determined from the g-dominated modes, whereas
the observable (p-dominated) modes have period differences that are lower
by a factor ranging from 1.2 to~2.

\begin{figure}
\centerline{\includegraphics[width=0.9\textwidth]{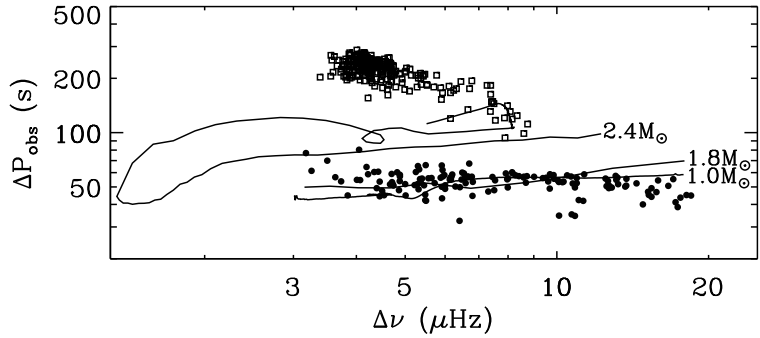}}
\caption{\label{fig.nature} Period spacings of $l=1$ modes in red giants.
  The symbols are measurements obtained with \kepler\
  \citep{kepler-bedding++2011-rg-nature}.  Filled circles indicate
  hydrogen-shell-burning giants and open squares indicate
  helium-core-burning stars.  The solid lines show model calculations using
  ASTEC and ADIPLS \citep{ChD2008a,ChD2008b}.  The period spacings were
  computed from the models by averaging over the observable modes
  \citep[see][for details]{white++2011-diagrams}.  The
  hydrogen-shell-burning giants evolve from right to left as they ascend
  the red giant branch.  The 2.4-$M_\odot$ model continues into the
  helium-core-burning stage, looping back to the right to join the
  so-called secondary clump \cite{Gir99}.  The lower-mass models stop
  before the helium-burning stage because they undergo a helium flash that
  cannot be calculated by ASTEC.  Solar metallicity was adopted for all
  models, which were computed without mass loss.  }
\end{figure}

Another way to visualise the frequencies is shown in Fig.~\ref{fig.rg}d.
This shows the model frequencies for $l=1$ in \'echelle format, but setting
the horizontal axis to be period modulo the g-mode period spacing (rather
than frequency modulo the large frequency separation).  Doing this for real
stars is difficult because most of the $l=1$ modes are not observable, so
that finding the correct period spacing is equivalent to estimating the
exact number of missing $l=1$ modes in each gap.
\citet{kepler-bedding++2011-rg-nature} were only able to do this
unambiguously for those stars with the largest number of detected modes.
For the remainder, however, it was still possible to measure the average
period spacing of the observed $l=1$ modes (\DPobs).  This quantity is
plotted in Fig.~\ref{fig.nature} against the large frequency separation,
and the stars are seen to divide nicely into two groups \citep[see
also][]{corot-mosser++2011-mixed}.  Theoretical models confirm that these
groups correspond to stars still burning hydrogen in a shell (filled
circles) and those also burning helium in the core (open squares; for more
details see \citealt{kepler-bedding++2011-rg-nature,white++2011-diagrams}).
These issues are discussed in detail in this volume by \citet{ChD2011}, who
shows that the larger period spacing in He-burning stars is due to the
presence of a convective core.

Finally, Simon O'Toole (priv. comm.) has pointed out that the close
agreement between the period spacings found in the He-burning red giants
with the those hot subdwarf B (sdB) stars from \kepler\ data
\citep{kepler-reed++2011-compact-VIII} is nicely consistent with the
accepted explanation that sdB stars are the cores of He-burning red giants
that have lost their envelopes.

\paragraph{Acknowledgments}
Many thanks to Pere Pall\'e and colleagues for organising this fantastic
Winter School.  I am grateful to Dennis Stello, Tim White and J{\o}rgen
Christensen-Dalsgaard for providing figures.  I also thank them, as well as
Hans Kjeldsen, Daniel Huber and Othman Benomar, for many helpful
discussions.

  \label{refs}
  \bibliographystyle{cambridgeauthordate}

\end{document}